      [1999/12/01 v1.4c Il Nuovo Cimento]
\documentclass{cimento}

\usepackage{graphicx}  
\usepackage{amssymb}
\title{Inclusion of $W^\pm$ single-spin asymmetry data\\
in a polarised PDF determination via Bayesian reweighting}
\shorttitle{Inclusion of $W^\pm$ single-spin asymmetry data
via Bayesian reweighting}
\author{E.~R.~Nocera\from{ins:x}}
\instlist{\inst{ins:x} Dipartimento di Fisica, Universit\`{a} di Milano and INFN, Sezione di Milano -\\ 
Via Celoria 16, I-20133 Milano, IT}
\PACSes{\PACSit{13.88.+e}{13.85.Ni}}
\begin{document}

\maketitle

\begin{abstract}
We discuss how the experimental 
information from longitudinal single-spin
asymmetries for $W^{\pm}$ boson production in polarised 
proton-proton collisions can be included 
in a polarised parton determination by Bayesian reweighting of a
Monte Carlo set of polarised PDF replicas.
We explicitly construct a prior ensemble of  
polarised parton distributions using available fits to
inclusive and semi-inclusive DIS data and we discuss the potential
impact of existing and future RHIC measurements on it.
\end{abstract}

Understanding the spin structure of the nucleon 
in terms of its parton substructure
remains a fundamental challenge in Quantum Chromodynamics (QCD).
In this framework, spin-dependent, or polarised, Parton Distribution
Functions (PDFs) play a leading role, since their first moments
are directly related to quark and gluon contributions to the total
nucleon spin (see for example~\cite{deFlorian:2011ia}
and references therein).

Several next-to-leading order (NLO) QCD analyses based on world data have been
carried out in those last years, aiming at an extraction of 
polarised PDF sets together with an accurate uncertainty 
estimation~\cite{Altarelli:1998nb,Blumlein:2010rn,Hirai:2008aj,Leader:2010rb,
deFlorian:2009vb}.
Different pieces of experimental data were included in these studies,
thus probing different aspects of the spin-dependent PDFs. 
Some analyses~\cite{Altarelli:1998nb,Blumlein:2010rn} 
were solely based on polarised inclusive deep-inelastic
lepton-nucleon scattering (DIS), which provides
information only on gluon and on the sum of quark and antiquark 
contributions.
Quark-antiquark separation
can be achieved by considering other processes, such as 
charged-current deep-inelastic 
scattering with neutrino beams (at a neutrino factory),
high-energy polarised proton-proton collisions with $W^{\pm}$
boson production at the Relativistic Heavy Ion Collider (RHIC) and
semi-inclusive deep-inelastic lepton-nucleon scattering (SIDIS).
Only data coming from SIDIS has been currently included
in more recent polarised global analyses~\cite{Leader:2010rb,deFlorian:2009vb}, 
but uncertainties on the polarised antiquark 
PDFs still remain relatively large.

Besides slight differences in the choice of datasets 
and in the details of the QCD analysis
(like treatment of higher twist effects), all these 
parton extractions are based on fixed functional 
forms for PDF parametrisation and on the Hessian approach 
for uncertainty estimation. This methodology is known 
to be affected by the intrinsic bias associated to the 
choice of some parametrisation, and by the limitations of linear
 error propagation. These issues result in a
systematic underestimation of uncertainties, 
especially in those kinematic regions where
the experimental constraints are loose. 
These drawbacks in the standard approach to
PDF determination are more severe for polarised PDFs,
due to the quantity and the quality of experimental data,
which are respectively less abundant and less accurate than 
their unpolarised counterparts.
 
The NNPDF Collaboration has developed an alternative unbiased methodology
for unpolarised parton fitting in 
recent years~\cite{DelDebbio:2007ee,Ball:2008by,
Ball:2009mk,Ball:2010de,Ball:2011uy}.
Monte Carlo sampling for error propagation
and Neural Networks used as unbiased interpolants are 
the main features of the NNPDF methodology, which provides 
a faithful statistical representation of PDFs and their uncertainties.
The NNPDF Collaboration regularly delivers updated sets of unpolarised 
PDFs, among which the first PDF set including all available
LHC data~\cite{Ball:2012cx}.

Preliminary results on the extraction of a polarised parton set,
based on the NNPDF methodology, suggest that some
polarised PDF uncertainties might be
underestimated in other available PDF determinations~\cite{Nocera:2012hx}.
This first NNPDF polarised set is based only on inclusive polarised DIS data,
hence it cannot disentangle quark and antiquark PDFs. A possible way
to achieve this result could be to include
SIDIS data, as in refs.~\cite{Leader:2010rb,
deFlorian:2009vb}. However, this requires the use of poorly known
polarised fragmentation functions, or else their simultaneous
determination 
with the NNPDF methodology.

The production of $W^{\pm}$ boson in (longitudinally) polarised 
proton-proton collisions at RHIC will provide independent and clean
access to the individual polarised quark and antiquark flavours
$\Delta u$, $\Delta\bar{u}$, $\Delta d$, $\Delta\bar{d}$~\cite{Bourrely:1993dd,Bunce:2000uv}.
The process is driven by purely weak interaction which 
couples left-handed quarks with right-handed antiquarks only
($u_L\bar{d}_R \rightarrow W^+$ and $d_L\bar{u}_R \rightarrow W^-$),
thus giving rise to a large $W$ parity-violating longitudinal single-spin
asymmetry which is sensitive to $\Delta q$ and $\Delta\bar{q}$ flavour dependence.
Production of $W$ bosons occurs at a scale where perturbative QCD is completely reliable
and it is free from fragmentation functions since 
$W$s are detected through leptonic decays.

The inclusion of RHIC $W$ boson production dataset in the 
polarised NNPDF parton fit~\cite{Nocera:2012hx} can be 
performed easily by means of the PDF reweighting 
technique described in refs.~\cite{Ball:2010gb,Ball:2011gg}. 
The method, based on statistical inference and Bayes theorem,
consists in assigning to each replica in a PDF ensemble 
a weight which assesses the probability that this replica 
agrees with the new data. These weights are computed by evaluating 
the $\chi^2$ of the new data to the prediction obtained 
using a given replica. The reweighted ensemble then
forms a representation of the probability distribution of 
PDFs conditional on both the original, or prior, ensemble 
(including old data) and the new data. 
It is worth noticing that, after the reweighting procedure,
replicas with small weights 
will become almost irrelevant in ensemble averages, 
thus being less efficient than their original 
counterparts at representing the underlying probability distribution.

A major problem in applying the reweighting technique to
RHIC data is  the 
fact that $W$ boson production
is sensitive to quark-antiquark separation, while
the prior polarised PDF ensemble, solely based on DIS data, only
contains the sum of quark and antiquark distributions for each flavour. We
must therefore construct a prior for the difference of these distributions.
This can be done using information on $\Delta\bar{u}$ and 
$\Delta\bar{d}$ coming from one of the aforementioned fits to SIDIS
data, such as
DSSV08 fit~\cite{deFlorian:2009vb}. 
If the new datasets bring in a sufficient amount of new information,
results will then be almost independent from the choice of the
prior~\cite{Giele:1998gw,Giele:2001mr}. 

More in detail, we proceed as follows. First of all,
we sample the $\Delta\bar{u}$ and $\Delta\bar{d}$ PDFs 
from DSSV08~\cite{deFlorian:2009vb} at a fixed reference
scale $Q_0^2=1$ \rm{GeV$^2$}. We select ten points, 
half logarithmically spaced and half linearly spaced 
in the range of momentum fraction
$10^{-3} \lesssim x \lesssim 0.4$, which roughly
corresponds to the interval of SIDIS experimental data relevant for 
disentangling quark-antiquark contributions.
Then, we generate $N_{\rm{rep}}=100$ replicas of these points
with a Gaussian distribution centered at the best fit and 
with standard deviation given by the DSSV08 PDF error estimate.
We check that average and variance computed on the $N_{\rm{rep}}$
replicas reproduce DSSV08 best fit and error for each sampled point
within percent accuracy. 
\begin{figure}[t]
\centering
\includegraphics[scale=0.3]{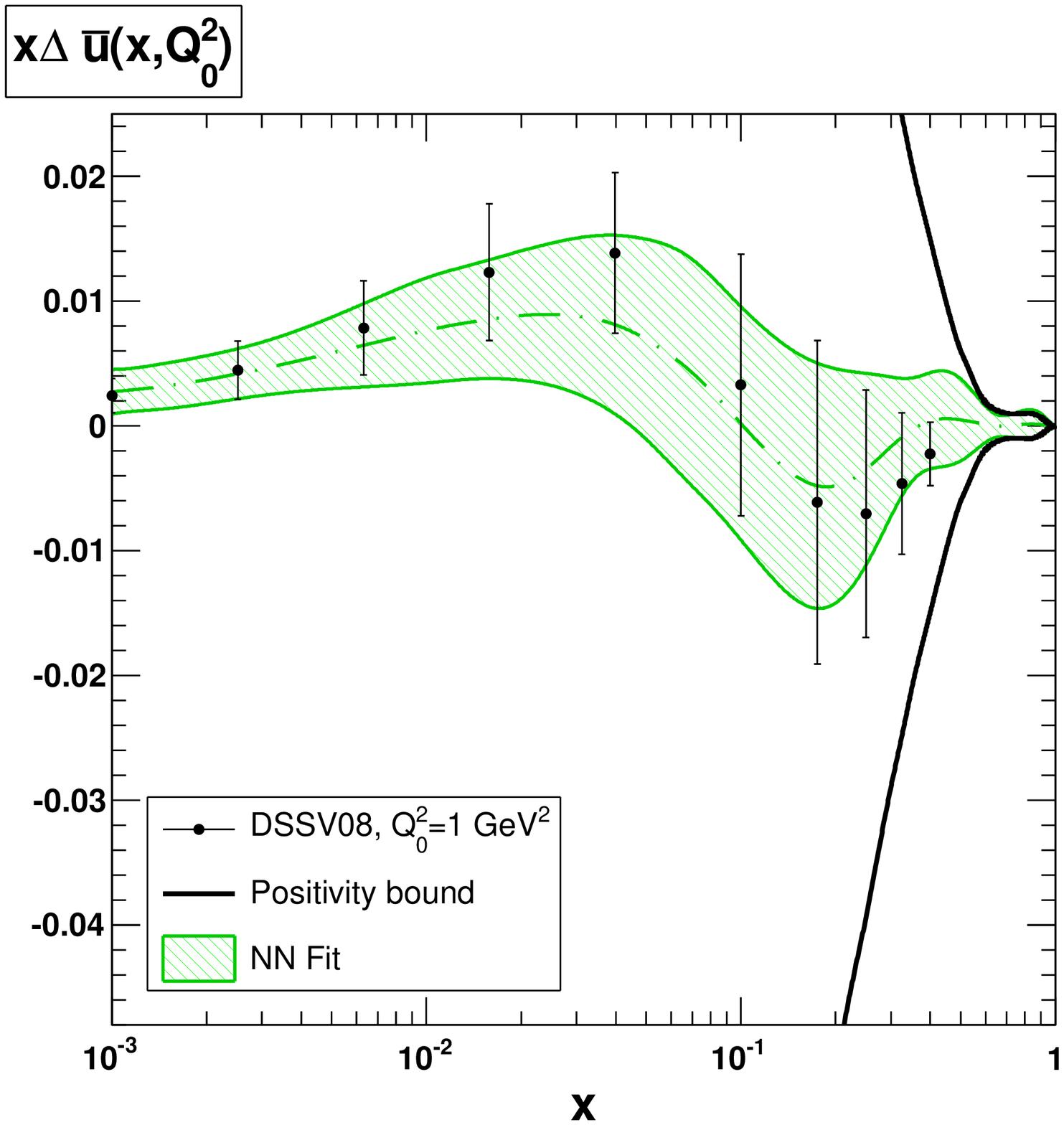}     
\includegraphics[scale=0.3]{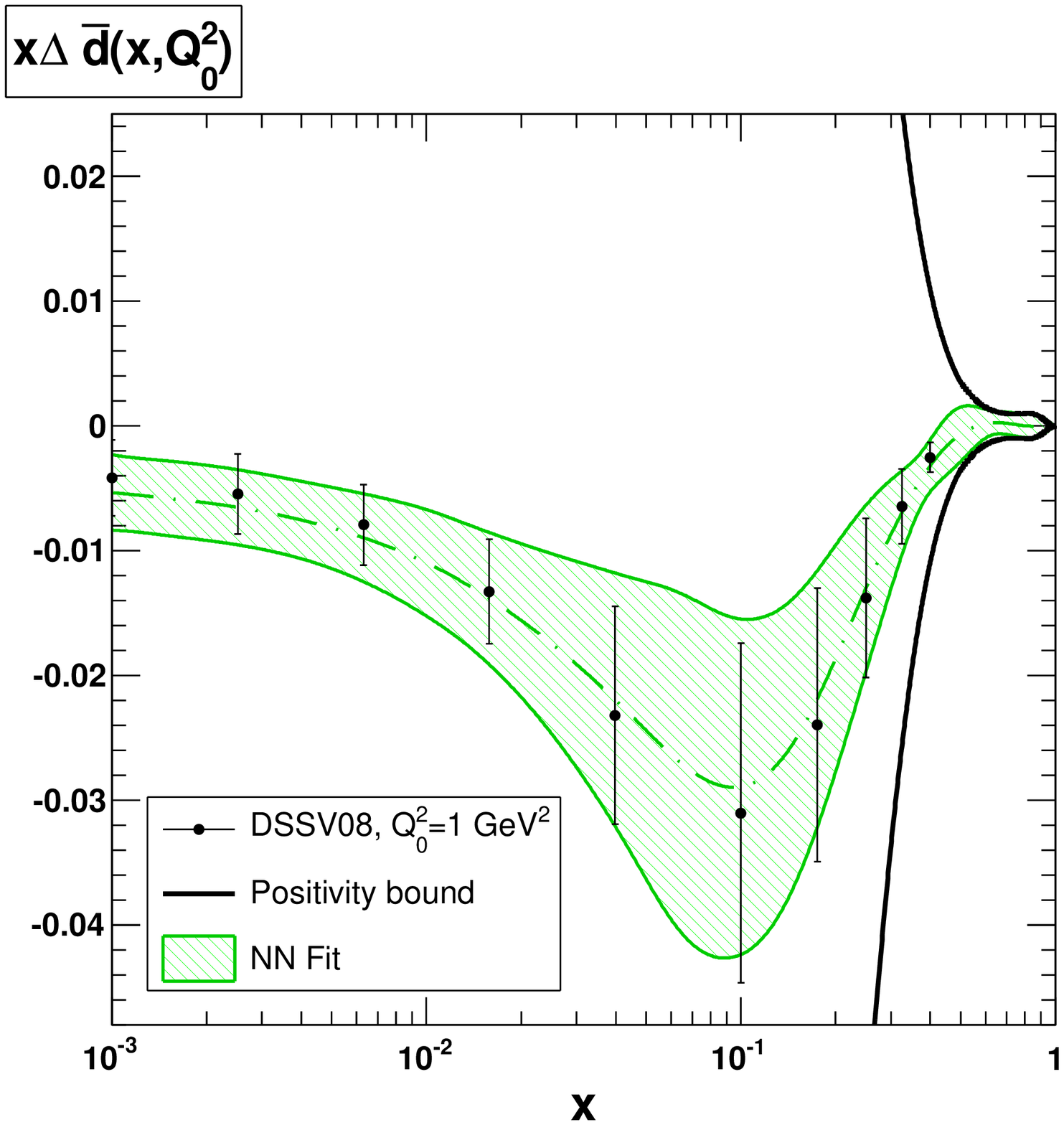}
\caption{The $x\Delta \bar{u}$ (left plot) and the $x\Delta \bar{d}$ (right plot) PDFs
from the neural network (NN) fit at the initial parametrisation scale
$Q_0^2=1$ \rm{GeV$^2$}. Data points sampled from the DSSV08
fit~\cite{deFlorian:2009vb} and the positivity bound (see the text) are also shown.}
\label{fig:ubardbar_fit}
\end{figure}

We supplement the input PDF basis in ref.~\cite{Nocera:2012hx}   
with two linear independent light quark-antiquark combinations, 
the total valence $\Delta V$ and the $\Delta V_8$,
which read, at the initial scale $Q_0^2$ and 
with the assumption $\Delta s = \Delta\bar{s}$,
\begin{eqnarray}
\Delta V(x,Q_0^2) &=& \Delta u^-(x,Q_0^2)+\Delta d^-(x,Q_0^2)\mbox{,} 
\\
\Delta V_8(x,Q_0^2) &=& \Delta u^-(x,Q_0^2)-\Delta d^-(x,Q_0^2)\mbox{,}
\label{eq:PDFbasis}
\end{eqnarray}
where $\Delta q^-=\Delta q-\Delta\bar{q}\mbox{, }q=u,d$.
Each of these PDFs is then parametrised by means of 
a neural network supplemented 
with a preprocessing polynomial and 
minimisation is performed using a genetic algorithm 
as in other NNPDF fits (for details on the general procedure 
see for example~\cite{Ball:2010de}).
Positivity constraints are taken into account during the fitting procedure
by penalising those replicas which do not fulfill the 
condition 
\begin{equation}
 |\Delta f(x, Q_0^2)| \leq f(x,Q_0^2) + \sigma(x,Q_0^2)\mbox{, } f=u,\bar{u},d,\bar{d}\mbox{, }
\end{equation}
where $f(x,Q_0^2)$ and $\sigma(x,Q_0^2)$ are the mean value and 
the standard deviation computed from the NNPDF2.1 NNLO parton fit~\cite{Ball:2011uy}.
We show the $x\Delta\bar{u}(x,Q_0^2)$ and $x\Delta\bar{d}(x,Q_0^2)$
PDFs resulting from our fit in fig.~\ref{fig:ubardbar_fit}.     

\begin{figure}[t]
\centering
\includegraphics[scale=0.3]{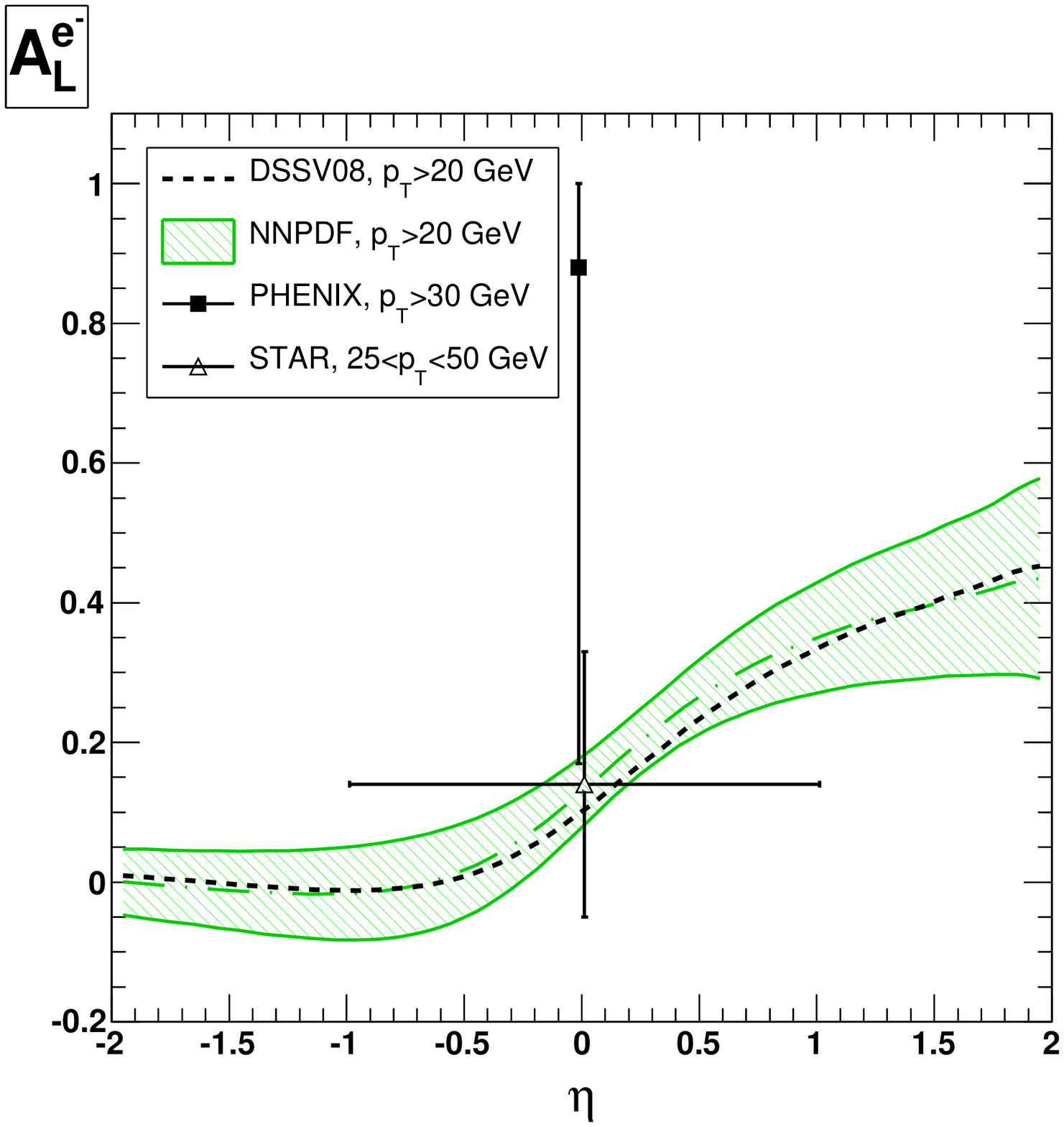}     
\includegraphics[scale=0.3]{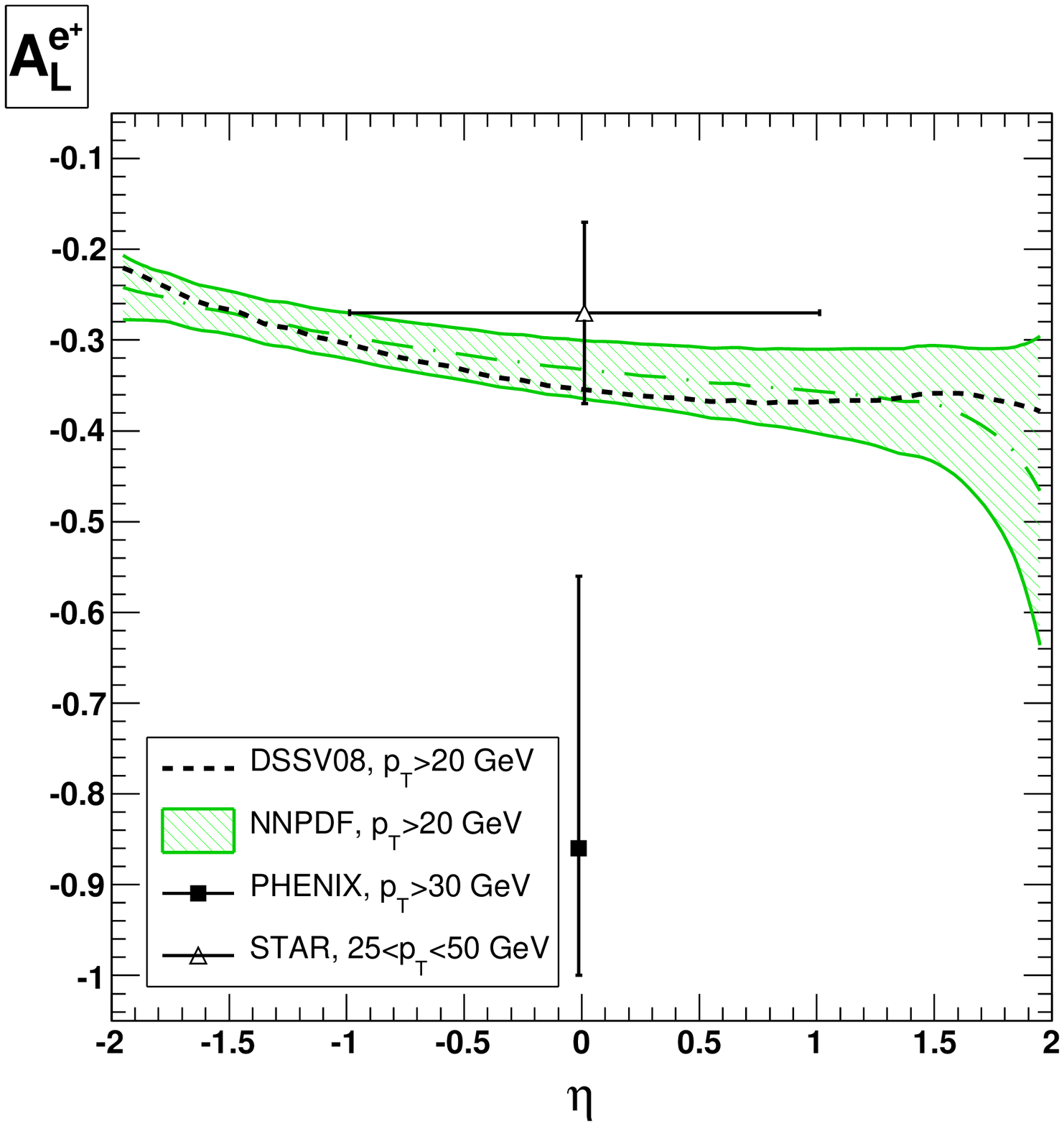}
\caption{Longitudinal single-spin asymmetry $A_L^{e^-}$ ($A_L^{e^+}$) 
for electron (positron) through production and decay 
of $W$ bosons at RHIC. The dashed curve corresponds
to the DSSV08 best fit PDFs, while the dashed-dotted curve 
and uncertainty band correspond to the mean value and one-sigma error 
computed within the present NNPDF analysis. We also show 
experimental measurements (run 2009) from PHENIX~\cite{Adare:2010xa} 
and STAR~\cite{Aggarwal:2010vc} collaborations
(uncertainties are statistical only).}
\label{fig:asy}
\end{figure}

We have used this prior polarised PDF ensemble, together with the
unpolarised NNPDF2.1 NLO parton set~\cite{Ball:2011uy}, to compute 
the observables measured by RHIC. To this purpose,
we have run the Monte-Carlo  code 
CHE~\cite{deFlorian:2010aa},
which we have modified to allow the usage of NNPDF ensembles.
In fig.~\ref{fig:asy}, we show the electron (positron)
longitudinal single-spin asymmetry $A_L^{e^-}$ ($A_L^{e^+}$) 
from  $W^{-(+)}$ boson production at RHIC computed at NLO using our
prior PDF set. 
The result is shown as a function of the lepton rapidity $\eta$,
at center-of-mass energy $\sqrt{s}=500$ \rm{GeV$^2$}. 
The experimental measurements from PHENIX~\cite{Adare:2010xa} 
and STAR~\cite{Aggarwal:2010vc} collaborations, and the DSSV central
prediction are also shown. All theoretical curves are obtained
integrating the 
lepton transverse momentum $p_T$  over all $p_T>20$ \rm{GeV$^2$},
while the data have $p_T < 30$~GeV (PHENIX) or $25 <p_T <50$~GeV (STAR).
Nevertheless, fig.~\ref{fig:asy}
interestingly sketches the impact of PDF uncertainties on
the asymmetry error estimate: at forward (backward) 
$\eta$ the uncertainty on $A_L^{e^{+(-)}}$ is large 
since it is correlated to the large uncertainty found for the $\bar{d}$ 
($\bar{u}$) polarised PDFs.
At mid-rapidity, $W^{+(-)}$ production probes a combination 
of the polarisation of the $u$ and $\bar{d}$ ($d$ and $\bar{u}$)
quarks, and $A_L^{e^{+(-)}}$ is expected to be negative 
(positive)~\cite{deFlorian:2010aa}. 
Clearly the data from the 2009 RHIC run shown in fig.~\ref{fig:asy}
will have little impact on the PDFs, but those from the 2012 run~\cite{Stevens:2012tc,Itaru:2012tc}
are likely to have a significant potential in disentangling the
individual polarised flavour and antiflavour distributions once
included via reweighting.

In summary, we have discussed how experimental data 
on longitudinal single-spin
asymmetries $A_L^{e^-}$ and $A_L^{e^+}$ 
from $W^{\pm}$ boson production at RHIC 
can be included in a 
polarised parton fit, based on the NNPDF methodology.
We have explicitly constructed a prior PDF ensemble
suited for reweighting with these datasets.
In future work, we will fully test the impact of RHIC data on 
our NNPDF polarised parton set through reweighting.
By varying the choice of prior 
PDF ensembles, for example by means of 
different assumptions on the antiquark distributions, we will then be
able to explicitly test for independence of results from the choice of prior,
and thus investigate the potential of RHIC data 
in providing insight into the nucleon spin structure. 

\acknowledgments
I would like to thank the organisers of
QCD-N'12 workshop for the opportunity to present this work
and D.~de~Florian for providing  the CHE code.
I am also grateful to  S.~Forte and  J.~Rojo
for useful discussions and for a careful reading of the 
manuscript.

\bibliography{nocera_QCDN12}
\bibliographystyle{varenna}

\end{document}